\documentclass[a4paper,11pt]{article}
\usepackage{graphicx}
\usepackage{amsfonts}
\usepackage{amssymb}
\usepackage{cite}

\begin{document}

\title{D-brane width}
\author{Youngsub Yoon}
\maketitle

\begin{abstract}

Loop quantum gravity predicts that there are non-zero minimal area, and non-zero minimal volume in (3+1) dimensions. Given this, one can easily guess that one will have a non-zero minimal 9-volume in (9+1) dimensions. Therefore, in this paper, we argue that not only D9-brane but also Dp-brane for p less than 9 has a 9-volume. This idea is new, as the present view states that such a Dp-brane has p-volume but no 9 volume. To demonstrate this, first, we equate D8-brane action with D9-brane action and show that 9th direction which is perpendicular to D8-brane has non-zero width. We repeat this step for different ps; we equate Dp-brane action with $D_{p-1}$ brane action. By this iteration and induction we conclude that Dp-brane has non-zero widths for each of (9-p) directions perpendicular to the Dp-brane, and therefore, non-zero 9 volume. When antisymmetric tensor and field strength are zero, this width is calculated to be $2 \pi \sqrt{\alpha'}$ for all (9-p) directions. For non-vanishing antisymmetric tensor and field strength, the width receives small corrections. In this paper, we only calculate up to the first order correction.
\end{abstract}
\pagebreak

The discovery of D-brane in mid 90s opened a new world for string theorists. [1] However, it was not known that all Dp-branes, regardless of p, have 9-volume, and therefore, non-zero widths, for p less than 9. This must be true, because one may easily guess that there is a non-zero minimal 9-volume in (9+1) dimensions, as there are non-zero minimal area and non-zero minimal volume in (3+1) dimensions according to loop quantum gravity. [2] Therefore, in this paper, we will argue that Dp-branes have 9-volume, and therefore have non-zero widths for p less than 9. This is in contrast with the present view that Dp-branes have zero 9 volume if p is less than 9. For example, according to the present view, a D3-brane may have 3-volume, but no 9-volume, because 6=9-3 spatial directions perpendicular to D3-brane is infinitely thin, or equal to zero length. However, as I will show Dp-brane can have widths for (9-p) directions perpendicular to the brane, it can have non-zero 9-volume.

To demonstrate this, we will first argue that the formula for D9-brane action can be used for any Dp-brane for arbitrary p, when suitably interpreted. This is justified, if one assumes that Dp-brane and D9-brane are made out of the same "matter," therefore, if D9-brane tension is used as universal tension for any p. This means that for vanishing antisymmetric tensor and field strength as D9-brane action is D9-brane tension multiplied by ten-dimensional world volume swiped by D9-brane, Dp-brane action is D9-brane tension multiplied by ten-dimensional world volume swiped by Dp-brane. To understand this explicitly, let's closely look at Dp-brane action which can be found in [3]. For the Dp-brane, the action is following.

\begin{equation}
S= T_{p}\, e^{-\phi} \int d^{p+1} \xi \,\sqrt{-det (G_{ab}+B_{ab}+2 \pi \alpha' F_{ab})}
\end{equation}
For p=9, we
\begin{equation}
S= T_{9}\, e^{-\phi} \int d^{10} \xi \,\sqrt{-det (G_{ab}+B_{ab}+2 \pi \alpha' F_{ab})}
\end{equation}
where a and b run from 0 to 9. For p=8 we get
\begin{equation}
S= T_{8}\, e^{-\phi} \int d^{9} \xi \,\sqrt{-det (G_{\mu\nu}+B_{\mu\nu}+2 \pi \alpha' F_{\mu\nu})}
\end{equation}
where $\mu$ and $\nu$ run from 0 to 8.
Now, let's use formula (2), a D9-brane action formula, to calculate D8-brane action, as claimed. First, note that
\begin{equation}
d^{10} x_{i}=d^{9} x_{i} d x_{9}
\end{equation}
Here, $x_{9}$ is the direction perpendicular to D8-brane. Plugging this relation to (2), we get
\begin{eqnarray}
S &=& T_{9}\, e^{-\phi} \int d^{9} x_{i} dx_{9} \sqrt{-det (G_{ab}+B_{ab}+2 \pi \alpha' F_{ab})} \nonumber\\
&=& T_{8} e^{-\phi} \int d^{9} x_{i} \sqrt{-det(G_{\mu\nu}+B_{\mu\nu}+ 2 \pi \alpha' F_{\mu\nu})}
\end{eqnarray}
To simplify this equation, note that,
\begin{eqnarray}
&& \sqrt{-det(G_{ab}+B_{ab}+ 2 \pi \alpha' F_{ab})} \nonumber\\
&=& \sqrt{-G_{99}\, det(G_{\mu\nu}+B_{\mu\nu}+2 \pi \alpha' F_{\mu\nu}) - \sum_{\mu= 1}^9 \frac{(B_{9\mu}+2 \pi \alpha' F_{9\mu})(B_{\mu9}+2 \pi \alpha' F_{\mu9})(det G_{ab})}{G_{\mu\mu} G_{99}}} \nonumber\\
&&
\end{eqnarray}
Here, we only considered up to the first order of antisymmetric tensor and field strength and assumed $G_{ab}$'s are diagonalized. Given this, by another one more approximation, we get the following.
\begin{eqnarray}
&& \frac{\sqrt{-det(G_{ab}+B_{ab}+ 2 \pi \alpha' F_{ab})}}{\sqrt{-det(G_{\mu\nu}+B_{\mu\nu}+2 \pi \alpha' F_{\mu\nu})}} \nonumber\\
&=& \sqrt{G_{99} - \sum_{\mu=1}^9 \frac{(B_{9\mu}+2 \pi \alpha' F_{9\mu})(B_{\mu9}+2 \pi \alpha' F_{\mu9})}{G_{\mu\mu}}}
\end{eqnarray}
Also note the following.
\begin{equation}
\int \sqrt{G_{99}}\, dx_{9} = w_9
\end{equation}
where $w_{9}$ is the width of D8-brane, as $\sqrt{G_{99}}$ is the correct Jacobian.
One more thing that we have to consider is the relationship between $T_{9}$ and $T_{8}$. A simple formula can be found in [3]. That is
\begin{equation}
T_{p}=\frac{T_{p-1}}{2 \pi \sqrt{\alpha'}}
\end{equation}
When p=9, we get
\begin{displaymath}
T_{9}=\frac{T_{8}}{2 \pi \sqrt{\alpha'}}
\end{displaymath}
Plugging (7),(8),(9) into (5) we get
\begin{eqnarray}
w_{9} &=& 2 \pi \frac{\sqrt{\alpha'}}{\sqrt{1+ (B_{\mu9}+ 2\pi \alpha' F_{\mu9})^2+O(F^{4})}} \nonumber\\
&=& 2 \pi \sqrt{\alpha'} \left( 1- \frac{1}{2}(B_{\mu9}+2 \pi \alpha' F_{\mu9})^{2}+O(F^{4}) \right)
\end{eqnarray}
Therefore, we got a formula for D8-brane width in the direction perpendicular to D8-brane. This whole step ((2) to (10)) can be done iteratively, for other values of p. For example, comparing D8-brane action and D7-brane action will give the width of D7-brane in the 8th direction, as comparing D9-brane action and D8-brane action gave the width of D8 brane in the 9th direction. Therefore, we arrive at the conclusion that Dp-brane has widths for (9-p) directions perpendicular to the Dp-brane. This width was calculated to be $2\pi \sqrt{\alpha'}$ for vanishing antisymmetric tensor and field strength and receives a small correction when they are non-vanishing. Also, notice that D-brane width is on the order of string scale.

In conclusion, D-branes cannot be infinitely thin; they have non-zero widths. Even D0-brane has non-zero widths, therefore it has a non-zero finite size or a 9-volume.

\pagebreak

\end{document}